%% file: minijl-ftfjp2019.tex
\documentclass[sigplan]{acmart}

\usepackage{booktabs}   
\usepackage{subcaption} 

\input{preamble}

\begin{document}

\title[Tag-Based Semantic Subtyping]{Decidable Tag-Based Semantic Subtyping \\
	for Nominal Types, Tuples, and Unions}


\author{Julia Belyakova}
\affiliation{
  \institution{Northeastern University}            
}
\email{belyakova.y@northeastern.edu}          


\begin{abstract}
Semantic subtyping enables simple, set-theoretical reasoning about types
by interpreting a type as the set of its values.
Previously, semantic subtyping has been studied primarily
in the context of statically typed languages with structural typing.
In this paper, we explore the applicability of semantic subtyping
in the context of a \emph{dynamic language with nominal} types.
Instead of static type checking, dynamic languages
rely on run-time checking of type tags associated with values,
so we propose using the tags for semantic subtyping.
We base our work on a fragment of the Julia language
and present \emph{tag-based semantic} subtyping 
for nominal types, tuples, and unions,
where types are interpreted set-theoretically
as sets of type tags. 
The proposed subtyping relation is shown to be decidable, 
and a corresponding analytic definition is provided.
The implications of using semantic subtyping
for multiple dispatch 
are also discussed.
\end{abstract}


\copyrightyear{2019}
\acmYear{2019}
\setcopyright{acmcopyright}
\acmConference[FTfJP'19]{Formal Techniques for Java-like Programs}{July 15, 2019}{London, United Kingdom}
\acmBooktitle{Formal Techniques for Java-like Programs (FTfJP'19), July 15, 2019, London, United Kingdom}
\acmPrice{15.00}
\acmDOI{10.1145/3340672.3341115}
\acmISBN{978-1-4503-6864-3/19/07}


\maketitle


\section{Introduction}\label{sec:intro}

\input{intro}

\section{Semantic Subtyping in \BetaJulia}\label{sec:semsub}

\input{types}

\section{Syntactic Definitions of Subtyping}\label{sec:synsub}

\input{declsub}

\input{redsub}

\section{Properties of Subtyping Relations}\label{sec:proofs}

\input{proofs}

\section{Semantic Subtyping and\\ Multiple Dynamic Dispatch}
\label{sec:discussion}

\input{discussion}

\section{Related Work}

\input{related-work}

\section{Conclusion and Future Work}

\input{conclusion}

\begin{acks}                            
   
   We are grateful to Ryan Culpepper, Artem Pelenitsyn, and Mitchell Wand 
   for insightful conversations.
   We thank Ellen Arteca, Benjamin Chung, Jane Kokernak, Artem Pelenitsyn, 
   Alexi Turcotte, Jan Vitek, and anonymous reviewers 
   for feedback on earlier drafts of the paper.
\end{acks}

\newpage

\bibliography{bibfile}

\clearpage

\appendix

\input{appendix}

\end{document}

%% file: preamble.tex
\usepackage{xspace}

\usepackage{xcolor}
\definecolor{light-gray}{gray}{0.85}

\usepackage[shortlabels]{enumitem}

\usepackage{mathpartir} 
\usepackage{mathtools} 

\usepackage{tikz}
\usetikzlibrary{fit,shapes, positioning}

\usepackage{listings}

\newcommand{\defemph}[1]{\textbf{#1}}

\newcommand{\R}[1]{\textsc{#1}\xspace}
\newcommand{\RD}[1]{\R{SD-#1}}
\newcommand{\RR}[1]{\R{SR-#1}}

\newcommand{\fn}[1]{\texttt{\small#1}}
\newcommand{\coqcode}[1]{\texttt{\small#1}}

\newcommand{\figref}[1]{Fig.~\ref{#1}}
\newcommand{\secref}[1]{Sec.~\ref{#1}}
\newcommand{\appref}[1]{App.~\ref{#1}}
\newcommand{\thmref}[1]{Theorem~\ref{#1}}

\newtheorem{theorem}{Theorem}

\newcommand{\BetaJulia}{\textsc{MiniJl}\xspace}

\newcommand{\jlcode}[1]{\texttt{\small#1}}

\newcommand{\jltype}[1]{\texttt{\small#1}}

\lstdefinelanguage[]{MiniJL}[]{Haskell}{
	morekeywords={Flt,Union},
	deletekeywords={print},
	escapeinside={\%*}{*)}, 
}

\lstnewenvironment{lstminijl}
{
	\lstset{
		language=MiniJl,
		basicstyle=\small\ttfamily,
		breaklines=true,
		columns=fullflexible
	}
}
{}

\newcommand{\Alt}{~\vert~}

\newcommand{\defsign}{
	\stackrel{\mathclap{\tiny\mbox{def}}}{\equiv}} 

\newcommand{\False}{\ensuremath{\mathtt{False}}\xspace}

\newcommand{\tyname}[1]{\ensuremath{\mathsf{#1}}}

\newcommand{\ty}{\ensuremath{\tau}\xspace}
\newcommand{\vty}{\ensuremath{v}\xspace}

\newcommand{\cname}{\ensuremath{\mathit{cname}}\xspace}
\newcommand{\aname}{\ensuremath{\mathit{aname}}\xspace}

\newcommand{\Type}{\textsc{Type}\xspace}
\newcommand{\VType}{\textsc{ValType}\xspace}
\newcommand{\PVType}{\ensuremath{\mathcal{P}(\VType)}}

\newcommand{\NomH}{\textrm{NomHrc}\xspace}

\newcommand{\semsubsign}{
	\stackrel{\mathclap{\tiny\mbox{sem}}}{<:}}

\newcommand{\tymtch}{\ensuremath{\ll}} 

\newcommand{\tyleq}{\ensuremath{\leq}}
\newcommand{\tyleqR}{\tyleq_\mathrm{R}}

\newcommand{\typair}[2]{\ensuremath{#1 \times #2}}

\newcommand{\tyunion}[2]{\ensuremath{#1 \cup #2}}

\newcommand{\tynum}{\tyname{Num}\xspace}
\newcommand{\tyreal}{\tyname{Real}\xspace}
\newcommand{\tyint}{\tyname{Int}\xspace}
\newcommand{\tyflt}{\tyname{Flt}\xspace}
\newcommand{\tycmplx}{\tyname{Cmplx}\xspace}
\newcommand{\tystr}{\tyname{Str}\xspace}

\newcommand{\tyref}{\tyname{Ref}\xspace}

\newcommand{\bjdeclsub}[2]{\ensuremath{#1 \rhd #2}}
\newcommand{\bjnomsub}[2]{\ensuremath{#1\,{\rhd^*} #2}}

\newcommand{\interpty}[1]{\ensuremath{\llbracket #1 \rrbracket}}

\newcommand{\bjmtch}[2]{\ensuremath{#1 \tymtch #2}}

\newcommand{\bjtruesemsub}[2]{\ensuremath{#1 \, \semsubsign \, #2}}
\newcommand{\bjsemsub}[2]{\ensuremath{#1 \, <: \, #2}}

\newcommand{\bjsub}[2]{\ensuremath{#1 \, \tyleq \, #2}}
\newcommand{\bjsubr}[2]{\ensuremath{#1 \, \tyleqR \, #2}}

\newcommand{\InNF}{\ensuremath{\mathop{\mathrm{InNF}}}\xspace}
\newcommand{\NF}{\ensuremath{\mathop{\mathrm{NF}}}\xspace}
\newcommand{\unprs}{\ensuremath{\mathop{\mathtt{un\_prs}}}\xspace}

\newcommand{\Atom}{\ensuremath{\mathop{\mathrm{Atom}}}\xspace}
\newcommand{\InNFAt}{\ensuremath{\mathop{\mathrm{InNF}_{\mathrm{at}}}}\xspace}
\newcommand{\NFAt}{\ensuremath{\mathop{\mathrm{NF}_{\mathrm{at}}}}\xspace}

%% file: intro.tex

In static type systems, subtyping is used to determine
when a value of one type can be safely used at another type.
It is often convenient to think of subtyping \jltype{T $<:$ S}
in terms of the set inclusion: ``the elements of~\jltype{T} are a subset
of the elements of~\jltype{S}''~\cite{bib:Pierce:2002:TAPL}.
This intuition is not always correct, but, in the case of
\emph{semantic subtyping}~\cite{bib:Hosoya:2003:XDuce,
	bib:Frisch:2008:sem-sub, bib:Ancona:2016:sem-sub-oo}, 
subtyping is defined exactly as the subset relation. 
Under semantic subtyping, types are interpreted as sets
$\interpty{\ty} = \{\nu \Alt \vdash \nu : \ty \}$, 
and subtyping $\ty_1 <: \ty_2$ is defined as inclusion 
of the interpretations
$\interpty{\ty_1} \subseteq \interpty{\ty_2}$.

Subtyping can also be used for run-time dispatch of function calls.
For example, object-oriented languages
usually support single dispatch~--- the ability to dispatch a method call 
based on the run-time type of the receiver object.
A more complex form of dispatch is \emph{multiple dispatch}
(MD)~\cite{bib:Chambers:1992:Cecil,bib:Clifton:2000:MultiJava},
which takes into account run-time types of \emph{all} arguments
when dispatching a function call.
One way to implement MD is to interpret both function signatures
and function calls
as tuple types~\cite{bib:Leavens:1998:mddtuples}
and then use subtyping on these types.

Dynamic dispatch is not limited to statically typed languages,
with multiple dispatch being even more widespread among 
\emph{dynamically} typed ones, e.g., CLOS, Julia, Clojure.
Unlike statically typed languages, 
which conservatively prevent type errors at compile-time, 
dynamic languages detect type errors at run-time:
whenever an operator is restricted to certain kinds of values,
the run-time system checks \emph{type tags} associated 
with the operator's arguments
to determine whether it can be safely executed.
A type tag indicates the run-time type of a value.
Thus, any class that can be instantiated induces a tag~--- 
the name of the class~--- 
whereas an abstract class or interface does not.
Some structural types also give rise to tags, 
e.g., tuples and sums (tagged unions).

While dynamically typed languages do use subtyping,
semantic subtyping is not applicable in this case,
for the semantic definition refers to a static typing relation.
To enable semantic reasoning in the context of dynamic languages,
we propose \emph{tag-based semantic} subtyping
where a type is interpreted as a set of run-time type tags 
instead of values. 

We define tag-based semantic subtyping for a fragment 
of the Julia language~\cite{bezanson2017julia}
that includes nominal types, tuples, and unions.
Tuples and unions are rather typical for semantic subtyping systems;
they have a clear set-theoretic interpretation 
and make up an expressive subtyping relation
where tuples distribute over unions.
At the same time, to the best of our knowledge,
the interaction of unions with \emph{nominal} types has not been studied before
in the context of semantic subtyping.
This interaction introduces
an unusual subtyping rule between abstract nominal types and unions,
with implications for multiple dispatch.
Note that the combination of unions and nominal types is not unique to Julia;
for instance, it also appears in the statically typed language 
Ceylon~\cite{bib:CeylonSpec:1:3}.

Our contributions are as follows:
\begin{enumerate}
  \item A definition of tag-based semantic subtyping for 
    nominal types, tuples, and unions (\secref{sec:semsub}).
  \item Two syntactic definitions of subtyping, 
    declarative (\secref{sec:declsub}) and reductive (\secref{sec:redsub}),
    along with Coq-mechanized proofs that these definitions are equivalent
    and coincide with the semantic definition (\secref{sec:proofs}). 	
  \item Proof of decidability of reductive subtyping (\appref{app:proofs}).
  \item Discussion of the implications of using semantic subtyping
    for multiple dispatch, as well as an alternative semantic interpretation
    of nominal types (\secref{sec:discussion}).
\end{enumerate}

%% file: types.tex
\begin{figure}
	\[
	\begin{array}{rcl@{\qquad}l}
	\ty \in \Type & ::= & & \text{\emph{Types}}
	\\ &\Alt& \typair{\ty_1}{\ty_2}  & \text{covariant pair}
	\\ &\Alt& \tyunion{\ty_1}{\ty_2} & \text{untagged union}
	\\ &\Alt& \cname  & \text{concrete nominal type}
	\\ &\Alt& \aname  & \text{abstract nominal type}
	\\ \\
	\cname & \in &
	\multicolumn{2}{l}{\{ \tyint, \tyflt, \tycmplx, \tystr \}}
	\\ 
	\aname & \in & \multicolumn{2}{l}{\{ \tyreal, \tynum \}}
	\end{array}
	\]
	\begin{tikzpicture}[sibling distance=4em, level distance=2.25em,
	concrete/.style = {shape=rectangle, draw, align=center}]
	\node { Num }
	child { node { Real }
		child { node[concrete] { Int} }
		child { node[concrete] (NF) { Flt} } }
	child { node[concrete, right=2em of NF] (NC) {Cmplx} }
	;
	\node[concrete, right=2em of NC] {Str} ;
	\end{tikzpicture}
	\caption{\BetaJulia: type grammar and nominal hierarchy}
	\label{fig:bjsem-types}
\end{figure}
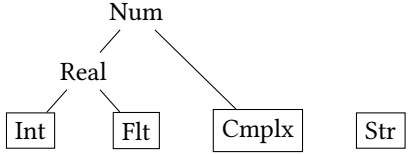

We base our work on a small language of types \BetaJulia,
presented in~\figref{fig:bjsem-types}.
Types, denoted by $\ty \in \Type$, include pairs, unions, 
and nominal types; \cname denotes \emph{concrete} nominal types
that can be instantiated,
and \aname denotes \emph{abstract} nominal types.

We work with a particular hierarchy of nominal types 
(presented in~\figref{fig:bjsem-types} as a tree)
instead of a generic class table to simplify the development.
There are four concrete leaf types (depicted in rectangles)
and two abstract types in the hierarchy. 
Formally, the hierarchy can be represented with a list of declarations
$\bjdeclsub{n_1}{n_2}$ read as ``$n_1$ extends $n_2$''
where $n$ is either $\cname$ or $\aname$.
In the case of \BetaJulia, the hierarchy is defined as follows:
\[
\NomH = [ \bjdeclsub{\tyreal}{\tynum}, 
\bjdeclsub{\tyint}{\tyreal}, \bjdeclsub{\tyflt}{\tyreal},
\bjdeclsub{\tycmplx}{\tynum} ].
\]
Nominal hierarchies should not have cycles, 
and each type can have only one parent.

\paragraph{Value Types}
Only instantiatable types induce type tags,
which we call \defemph{value types}. 
Their formal definition is given in~\figref{fig:bjsem-value-types}:
value type $\vty \in \VType$ is either a concrete nominal type 
or a pair of value types. 
For example, \tyflt, \typair{\tyint}{\tyint},
and \typair{\tystr}{(\typair{\tyint}{\tyint})} are all value types.
Union types, like abstract nominal types, are not value types.
Therefore, a type such as \tyunion{\tyint}{\tyint} is not a value type 
despite it describing the same set of values as the value type \tyint.

\begin{figure} 
	\[
	\begin{array}{rcl@{\qquad}l}
	\vty \in \VType & ::= & & \text{\emph{Value Types}}
	\\ &\Alt& \cname & \text{concrete nominal type}
	\\ &\Alt& \typair{\vty_1}{\vty_2} & \text{pair of value types}
	\end{array}
	\]
	\caption{Value types} 
	\label{fig:bjsem-value-types}
\end{figure}

\subsection{Semantic Interpretation of Types}\label{subsec:interp}

As mentioned in~\secref{sec:intro}, 
we interpret a type as a set of type tags (i.e. value types) instead of values
and call this semantic interpretation \emph{tag-based}.
Formally, the interpretation is given by the function
$\interpty{\cdot}$ that maps a type $\ty \in \Type$
into a set of value types $s \in \PVType$,
as presented in~\figref{fig:bjsem-interpretation}.

\begin{figure} 
	\[
	\begin{array}{rcl}
	\interpty{\cdot}: \Type &\rightarrow& \PVType \\
	\interpty{\cname}  & = & \{\cname\} \\
	\interpty{\tyreal} & = & \{ \tyint, \tyflt \} \\
	\interpty{\tynum} & = & \{ \tyint, \tyflt, \tycmplx \} \\
	\interpty{\typair{\ty_1}{\ty_2}} & = & \{\typair{\vty_1}{\vty_2} 
	\Alt \vty_1 \in \interpty{\ty_1}, \vty_2 \in \interpty{\ty_2}\}\\
	\interpty{\tyunion{\ty_1}{\ty_2}} & = & 
	\interpty{\ty_1} \cup \interpty{\ty_2}
	\end{array}
	\]
	\caption{Tag-based semantic interpretation of types} 
	\label{fig:bjsem-interpretation}
\end{figure}

A type's interpretation states what values constitute the type:
$\vty \in \interpty{\ty}$ means that values $\nu$ tagged with \vty
(i.e. instances of \vty) belong to $\ty$.
Thus, in \BetaJulia, a \emph{concrete nominal} type \cname is comprised 
only of its direct instances.\footnote{In the general case, the interpretation
of a concrete nominal type would include the type and all its concrete subtypes.}
\emph{Abstract nominal} types cannot be instantiated, 
but their interpretation needs to reflect the nominal hierarchy.
For example, a \tynum value 
is either a concrete complex or real number, which in turn
is either a concrete integer or a floating point value.
Therefore, the set of value types $\{\tycmplx, \tyint, \tyflt\}$
describes the set of all possible values of type \tynum.
More generally, the interpretation of an abstract nominal type \aname
can be given as follows: 
\[
\interpty{\aname} = \{\cname\ \Alt \bjnomsub{\cname}{\aname} \},
\]
where the relation $\bjnomsub{n_1}{n_2}$ means that 
nominal type $n_1$ transitively extends~$n_2$:
\begin{mathpar}
	\inferrule*[right=]
	{ \bjdeclsub{n_1}{n_2} \in \NomH }
	{ \bjnomsub{n_1}{n_2} }
	
	\inferrule*[right=]
	{ \bjnomsub{n_1}{n_2} \\ \bjnomsub{n_2}{n_3} }
	{ \bjnomsub{n_1}{n_3} }.
\end{mathpar}
Finally, \emph{pairs} and \emph{unions} are interpreted set-theoretically
as in standard semantic subtyping.

Once we have the tag interpretation of types, we define 
\defemph{tag-based semantic subtyping} in the usual manner~--- 
as the subset relation:
\begin{equation}\label{eq:truesemsub-def}
\bjtruesemsub{\ty_1}{\ty_2} \quad \defsign \quad
\interpty{\ty_1} \subseteq \interpty{\ty_2}.
\end{equation}


%% file: declsub.tex
While the semantic approach does enable intuitive set-theo\-re\-tic
reasoning about subtyping,
a subtyping relation also needs to be computable.
However, the semantic definition~\eqref{eq:truesemsub-def}
does not suit this purpose,
as it operates on interpretations.
In the general case, the interpretation of a type 
can be an infinite set, and as such, it cannot be computed.
In the finite case, generating the interpretation sets
and checking the subset relation on them would be inefficient.
Therefore, we provide an alternative, \emph{syntactic} definition of subtyping 
that is equivalent to~\eqref{eq:truesemsub-def}
and straightforward to implement.

We do this in two steps. 
First, we give an inductive \emph{declarative} definition
that is handy to reason about 
and prove it equivalent to the semantic definition.
Second, we provide a \emph{reductive} analytic\footnote{Inference rules
	are called \emph{analytic}~\cite{bib:Martin-Lof1994} 
	if there is a finite number of rules applicable to a judgment,
    and the premises of each rule are comprised of the subcomponents 
	of its conclusion. Such rules give rise to a straightforward 
	bottom-up algorithm.
	If there is always only one rule applicable to a judgment,
	analytic rules are called \emph{syntax-directed}.}
definition of subtyping
and prove it equivalent to the declarative one 
(and, hence, the semantic definition as well).
We prove that the reductive subtyping relation is decidable, 
i.e. for any two types $\ty_1$ and $\ty_2$,
it is possible to prove that either $\ty_1$ is a subtype of $\ty_2$ 
or it is not.
The proofs are mechanized in Coq, and since Coq logic is constructive,
the decidability proof is also a subtyping algorithm.
The algorithm can also be implemented as a straightforward recursive function.

\subsection{Declarative Subtyping}\label{sec:declsub}

The declarative syntactic definition of subtyping is provided in~\figref{fig:bjsem-decl-sub}.
It comprises most of the standard rules
of syntactic subtyping for unions and pairs:
reflexivity and transitivity (\RD{Refl} and \RD{Trans}), 
subtyping of pairs (\RD{Pairs}),
and subtyping of unions (\RD{UnionL}, \RD{UnionR1}, \RD{UnionR2}).
Though \RD{UnionR*} rules are seemingly very strict 
(they require the left-hand side type to be syntactically equivalent
to a part of the right-hand side type), 
transitivity allows us to derive judgments such as
$\bjsub{\tyint}{(\tyunion{\tystr}{\tyreal})}$ via
$\bjsub{\tyint}{\tyreal}$ and $\bjsub{\tyreal}{\tyunion{\tystr}{\tyreal}}$.

Note that all rules from~\figref{fig:bjsem-decl-sub} \emph{are essential} 
for the definition to be equivalent to semantic subtyping.
Thus, for example, the syntactic definition needs to be
{reflexive} and {transitive} because so is the subset relation,
which is used to define semantic subtyping.
Semantic subtyping also forces us to add rules 
for distributing pairs over unions, \RD{Distr1} and \RD{Distr2}. 
For instance, consider two types,
\typair{\tystr}{(\tyunion{\tyint}{\tyflt})}
and \tyunion{(\typair{\tystr}{\tyint})}{(\typair{\tystr}{\tyflt})}.
They have the same semantic interpretation~---
$\{\typair{\tystr}{\tyint}, \typair{\tystr}{\tyflt}\}$~---
so they are equivalent.
Therefore, we should also be able to derive their equivalence
using the declarative definition,
i.e. declarative subtyping should hold in both directions.
One direction is trivial:
\begin{mathpar}{\small
\inferrule*[right=]
{ \inferrule*[right=]
  { \bjsub{\tystr}{\tystr} \\ \bjsub{\tyint}{\tyunion{\tyint}{\tyflt}} }
  { \bjsub{\typair{\tystr}{\tyint}}
  	  {\typair{\tystr}{(\tyunion{\tyint}{\tyflt})}} } \\
  \inferrule*[right=]
  { \ldots }
  { \bjsub{\typair{\tystr}{\tyflt}}
  	  {\ldots} } }
{ \bjsub{\tyunion{(\typair{\tystr}{\tyint})}{(\typair{\tystr}{\tyflt})}}
	{\typair{\tystr}{(\tyunion{\tyint}{\tyflt})}} }.
}\end{mathpar}
But the other direction,  
\[
\bjsub{\typair{\tystr}{(\tyunion{\tyint}{\tyflt})}}
  {\tyunion{(\typair{\tystr}{\tyint})}{(\typair{\tystr}{\tyflt})}},
\]
cannot be derived without \RD{Distr2} rule. 

The novel part of the definition resides in subtyping of nominal types.
There are four obvious rules coming directly 
from the nominal hierarchy, for instance, \RD{RealNum} mirrors the fact 
that $\bjdeclsub{\tyreal}{\tynum} \in \NomH$.
But the rules \RD{RealUnion} and \RD{NumUnion}
(\colorbox{light-gray}{highlighted} in~\figref{fig:bjsem-decl-sub})
are new, dictated by semantic subtyping.
Thus, \RD{RealUnion} allows us to prove the equivalence
of types \tyunion{\tyint}{\tyflt} and \tyreal, 
which are both interpreted as $\{\tyint, \tyflt\}$.



\begin{figure}
	\begin{mathpar}
		\fbox{$\bjsub{\ty}{\ty'}$}\\
		
		\inferrule*[right=SD-Refl]
		{ }
		{ \bjsub{\ty}{\ty} }
		
		\inferrule*[right=SD-Trans]
		{ \bjsub{\ty_1}{\ty_2} \\ \bjsub{\ty_2}{\ty_3} }
		{ \bjsub{\ty_1}{\ty_3} }		
		\\
		
		\inferrule[SD-IntReal]
		{ }
		{ \bjsub{\tyint}{\tyreal} }
		
		\inferrule[SD-FltReal]
		{ }
		{ \bjsub{\tyflt}{\tyreal} }
		\\
		
		\inferrule[{SD-RealNum}]
		{ }
		{ \bjsub{\tyreal}{\tynum} }
	
		\inferrule[{SD-CmplxNum}]
		{ }
		{ \bjsub{\tycmplx}{\tynum} }
		\\
		
		\colorbox{light-gray}{$
		\inferrule[SD-RealUnion]
		{ }
		{ \bjsub{\tyreal}{\tyunion{\tyint}{\tyflt}} }
		$}
		
		\colorbox{light-gray}{$
		\inferrule[SD-NumUnion]
		{ }
		{ \bjsub{\tynum}{\tyunion{\tyreal}{\tycmplx}} }
		$}
		\\
		
		\inferrule*[right=SD-Pair]
		{ \bjsub{\ty_1}{\ty'_1} \\ \bjsub{\ty_2}{\ty'_2} }
		{ \bjsub{\typair{\ty_1}{\ty_2}}{\typair{\ty'_1}{\ty'_2}} }
		\\
		
		\inferrule[SD-UnionL]
		{ \bjsub{\ty_1}{\ty'} \\ \bjsub{\ty_2}{\ty'} }
		{ \bjsub{\tyunion{\ty_1}{\ty_2}}{\ty'} }
		
		\inferrule[{SD-UnionR1}]
		{ }
		{ \bjsub{\ty_1}{\tyunion{\ty_1}{\ty_2}} }
		
		\inferrule[{SD-UnionR2}]
		{ }
		{ \bjsub{\ty_2}{\tyunion{\ty_1}{\ty_2}} }
		\\
		
		\inferrule*[right=SD-Distr1]
		{ }
		{ \bjsub{\typair{(\tyunion{\ty_{11}}{\ty_{12}})}{\ty_2}}
			{\tyunion{(\typair{\ty_{11}}{\ty_2})}{(\typair{\ty_{12}}{\ty_2})}} }
		
		\inferrule*[right=SD-Distr2]
		{ }
		{ \bjsub{\typair{\ty_1}{(\tyunion{\ty_{21}}{\ty_{22}})}}
			{\tyunion{(\typair{\ty_1}{\ty_{21}})}{(\typair{\ty_1}{\ty_{22}})}} }
	\end{mathpar}
	\caption{Declarative subtyping for \BetaJulia}
	\label{fig:bjsem-decl-sub}
\end{figure}

%% file: redsub.tex
\subsection{Reductive Subtyping}\label{sec:redsub}

The declarative definition is neither syntax-directed nor analytic
and cannot be directly turned into a subtyping algorithm.
For one, the transitivity rule \RD{Trans} 
overlaps with any other rule in the system
and also requires ``coming up'' with an intermediate type $\ty_2$
to conclude $\bjsub{\ty_1}{\ty_3}$.
For instance, to derive 
\[\bjsub{\typair{\tystr}{\tyreal}}
{(\typair{\tystr}{\tyint}) \cup (\typair{\tystr}{\tystr}) 
	\cup (\typair{\tystr}{\tyflt})},\]
we need to apply transitivity several times, in particular, 
with the intermediate type $\typair{\tystr}{(\tyunion{\tyint}{\tyflt})}$.
Another source of overlap is the reflexivity and distributivity rules.

\begin{figure}
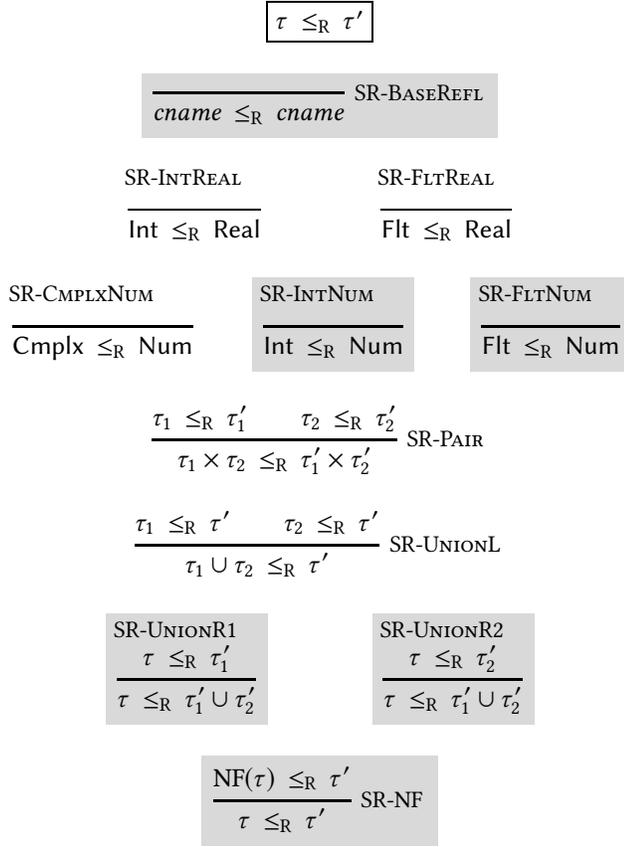

	\begin{mathpar}
		\fbox{$\bjsubr{\ty}{\ty'}$}\\
		
		\colorbox{light-gray}{$
		\inferrule*[right=SR-BaseRefl]
		{ }
		{ \bjsubr{\cname}{\cname} }
		$}
		\\
		
		\inferrule[SR-IntReal]
		{ }
		{ \bjsubr{\tyint}{\tyreal} }
		
		\inferrule[SR-FltReal]
		{ }
		{ \bjsubr{\tyflt}{\tyreal} }
		\\
	
		\inferrule[SR-CmplxNum]
		{ }
		{ \bjsubr{\tycmplx}{\tynum} }
		
		\colorbox{light-gray}{$
		\inferrule[SR-IntNum]
		{ }
		{ \bjsubr{\tyint}{\tynum} }
		$}
		
		\colorbox{light-gray}{$
		\inferrule[SR-FltNum]
		{ }
		{ \bjsubr{\tyflt}{\tynum} }
		$}
		\\
		
		\inferrule*[right=SR-Pair]
		{ \bjsubr{\ty_1}{\ty'_1} \\ \bjsubr{\ty_2}{\ty'_2} }
		{ \bjsubr{\typair{\ty_1}{\ty_2}}{\typair{\ty'_1}{\ty'_2}} }
		\\
		
		\inferrule*[right=SR-UnionL]
		{ \bjsubr{\ty_1}{\ty'} \\ \bjsubr{\ty_2}{\ty'} }
		{ \bjsubr{\tyunion{\ty_1}{\ty_2}}{\ty'} }
		\\
		
		\colorbox{light-gray}{$
		\inferrule[SR-UnionR1]
		{ \bjsubr{\ty}{\ty'_1} }
		{ \bjsubr{\ty}{\tyunion{\ty'_1}{\ty'_2}} }
		$}
		
		\colorbox{light-gray}{$
		\inferrule[SR-UnionR2]
		{ \bjsubr{\ty}{\ty'_2} }
		{ \bjsubr{\ty}{\tyunion{\ty'_1}{\ty'_2}} }
		$}
		\\
		
		\colorbox{light-gray}{$
		\inferrule*[right=SR-NF]
		{ \bjsubr{\NF(\ty)}{\ty'} }
		{ \bjsubr{\ty}{\ty'} }
		$}
	\end{mathpar}
	\caption{Reductive subtyping for \BetaJulia}
	\label{fig:bjsem-red-sub}
\end{figure}

By contrast, the rules of reductive subtyping enable
straightforward bottom up reasoning;
the rules are presented in~\figref{fig:bjsem-red-sub}.
The reductive definition lacks the most problematic rules
of declarative subtyping, 
i.e. general reflexivity, transitivity, and distributivity.
Some of the inductive rules have the exact declarative counterparts,
e.g. subtyping of pairs (\RR{Pair}) or
subtyping of a union on the left (\RR{UnionL}).

The differing rules are \colorbox{light-gray}{highlighted}.
The explicit reflexivity rule \RR{BaseRefl} now only works with 
concrete nominal types, 
but this already makes the reductive definition reflexive.
The definition also has to be transitive,
so several rules are added or modified to enable derivations
that used to rely on transitivity in the declarative definition.
These include subtyping of nominal types (\RR{IntNum}, \RR{FltNum}),
subtyping of a union on the right (\RR{UnionR1}, \RR{UnionR2}),
and normalization (\RR{NF}).

\begin{figure}
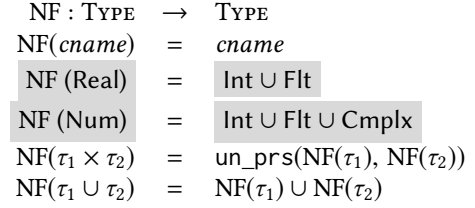

  \[
	\begin{array}{rcl}
	\NF: \Type &\rightarrow& \Type \\
	\NF(\cname) &=& \cname \\
	\colorbox{light-gray}{\NF(\tyreal)} &=&
	\colorbox{light-gray}{\tyunion{\tyint}{\tyflt}} \\
	\colorbox{light-gray}{\NF(\tynum)} &=&
	\colorbox{light-gray}{\tyunion{\tyunion{\tyint}{\tyflt}}{\tycmplx}} \\
	\NF(\typair{\ty_1}{\ty_2}) &=& \unprs(\NF(\ty_1), \, \NF(\ty_2))	\\
	\NF(\tyunion{\ty_1}{\ty_2}) &=& \tyunion{\NF(\ty_1)}{\NF(\ty_2)} \\
	\end{array}
  \]
	\caption{Computing normal form of \BetaJulia types}
	\label{fig:bjsem-calc-nf}
\end{figure}

The last rule of the definition, \RR{NF}, is the most important,
as it covers all useful interactions of transitivity and distributivity 
that are possible in the declarative definition.
The rule rewrites type \ty into its \defemph{normal form} $\NF(\ty)$
before applying other subtyping rules.
Any normalized type has the form $\vty_1 \cup \vty_2 \cup \ldots \cup \vty_n$,
i.e. a union of value types
(we omit parenthesis because union is associative).
The normalization function $\NF$ is presented in~\figref{fig:bjsem-calc-nf}
(the auxiliary function $\unprs$ 
can be found in~\figref{fig:bjsem-calc-nf-full}, \appref{app:nf}).
It produces a type in \emph{disjunctive normal form}
by replacing an abstract nominal type 
with the union of all its concrete subtypes, 
and a pair of unions with the union of pairs of value types
(each of this pairs is itself a value type),
for instance:
\[
\NF(\typair{\tystr}{(\tyunion{\tyint}{\tyflt})}) =
\tyunion{(\typair{\tystr}{\tyint})}{(\typair{\tystr}{\tyflt})}.
\]
As shown in~\secref{sec:declsub-correct}, a type and its normal form are
equivalent according to the declarative definition.
This property is essential for the reductive subtyping 
being equivalent to the declarative one.

\paragraph{Subtyping Algorithm.}
The reductive rules are analytic,
and if a derivation of $\bjsub{\ty}{\ty'}$ exists,
it can always be found by the following algorithm.
\begin{enumerate}
  \item Use the normalization rule \RR{NF} once (normalize $\ty$);
  \item Use all the other rules to derive 
    $\bjsub{\NF(\ty)}{\ty'}$ in the standard manner, bottom up;
    except for an overlap between \RR{UnionR1} and \RR{UnionR2},
    these rules are syntax-directed.
\end{enumerate}

However, this algorithm does not always produce the shortest derivation.
For instance, for
$\bjsubr{\typair{\tystr}{(\tyunion{\tyint}{\tyflt})}}
	    {\typair{\tystr}{\tyreal}}$,
it produces a derivation with eight applications of the rules, 
whereas the shortest derivation needs only five applications
(see~\appref{app:example-deriv}).
It is possible that in practice, an algorithm that tries the short path first 
and only then resorts to normalization would work better.

The actual Julia implementation uses a clever algorithm 
to check subtyping of tuples and unions 
without having to normalize types~\cite{bib:Chung19}.
The algorithm is equivalent to the normalization-based one discussed above,
but instead of computing the whole normal form, 
it computes only the components of the normalized type, one at a time.

Note that the rules for subtyping of nominal types do not have to be built-in.
Instead of five separate rules, as presented in~\figref{fig:bjsem-red-sub},
we can use a single rule that relies on the relation 
$\bjnomsub{n_1}{n_2}$ ($n_1$ transitively extends $n_2$)
from~\secref{subsec:interp}:
\begin{mathpar}
	\inferrule*[right=SR-Nom]
	{ \bjnomsub{n_1}{n_2} }
	{ \bjsub{n_1}{n_2} }.
\end{mathpar}
Then, for any $n_1$ and $n_2$, the relation $\bjnomsub{n_1}{n_2}$ 
can be checked algorithmically, using the nominal hierarchy $\NomH$.



%% file: proofs.tex
\subsection{Correctness of Declarative Subtyping}\label{sec:declsub-correct}

\begin{figure}
	\begin{mathpar}
		\fbox{$\bjmtch{\ty}{\ty'}$}\\
		
		\inferrule*[right=MT-CName]
		{ }
		{ \bjmtch{\cname}{\cname} }		
		\\
		
		\inferrule[MT-IntReal]
		{ }
		{ \bjmtch{\tyint}{\tyreal} }
		
		\inferrule[MT-FltReal]
		{ }
		{ \bjmtch{\tyflt}{\tyreal} }
		\\
		
		\inferrule[MT-IntNum]
		{ }
		{ \bjmtch{\tyint}{\tynum} }
		
		\inferrule[MT-FltNum]
		{ }
		{ \bjmtch{\tyflt}{\tynum} }
		
		\inferrule[MT-CmplxNum]
		{ }
		{ \bjmtch{\tycmplx}{\tynum} }
		\\
		
		\inferrule*[right=MT-Pair]
		{ \bjmtch{\vty_1}{\ty_1} \\ \bjmtch{\vty_2}{\ty_2} }
		{ \bjmtch{\typair{\vty_1}{\vty_2}}{\typair{\ty_1}{\ty_2}} }
		\\
		
		\inferrule*[right=MT-Union1]
		{ \bjmtch{\vty}{\ty_1}  }
		{ \bjmtch{\vty}{\tyunion{\ty_1}{\ty_2}} }
		
		\inferrule*[right=MT-Union2]
		{ \bjmtch{\vty}{\ty_2}  }
		{ \bjmtch{\vty}{\tyunion{\ty_1}{\ty_2}} }
	\end{mathpar}
	\caption{Matching relation in \BetaJulia}
	\label{fig:bjsem-match}
\end{figure}

In order to show correctness of declarative subtyping,
we need to prove that the declarative definition of subtyping 
is sound and complete with respect to the semantic definition.
Formally, we write this statement as:
\begin{equation}\label{eq:declsub-correct-truesemsub}
\forall \ty_1, \ty_2.\ (\bjsub{\ty_1}{\ty_2} \iff \bjtruesemsub{\ty_1}{\ty_2}).
\end{equation}

Instead of directly proving~\eqref{eq:declsub-correct-truesemsub},
it is more convenient to prove the equivalence of declarative subtyping
to the following relation
(referred to as \defemph{matching-based semantic subtyping}):
\begin{equation}\label{eq:semsub-def}
\bjsemsub{\ty_1}{\ty_2} \quad \defsign \quad
\forall \vty.\ (\bjmtch{\vty}{\ty_1} \implies\ \bjmtch{\vty}{\ty_2}).
\end{equation}
The definition~\eqref{eq:semsub-def}
relies on the relation $\bjmtch{\vty}{\ty}$ 
(defined in~\figref{fig:bjsem-match}), read ``tag~\vty matches type~\ty'', 
which we call the \defemph{matching relation}.

Tag-based and matching-based semantic subtyping relations 
are equivalent:
\[
\forall \ty_1, \ty_2.\ 
(\bjsemsub{\ty_1}{\ty_2} \iff \bjtruesemsub{\ty_1}{\ty_2}).
\]
To see why, let us recall that tag-based semantic subtyping~\eqref{eq:truesemsub-def} 
is defined as $\interpty{\ty_1} \subseteq \interpty{\ty_2}$
and the subset relation $X \subseteq Y$ as
$\forall x.\ (x \in X \implies x \in Y)$.
Therefore, the definition~\eqref{eq:truesemsub-def} can be 
rewritten as:
\begin{equation}\label{eq:truesemsub-def-new}
\bjtruesemsub{\ty_1}{\ty_2} \;\defsign \;
\forall \vty.\ (\vty \in \interpty{\ty_1} \implies \vty \in \interpty{\ty_2}).
\end{equation}
It is easy to show by induction on \ty that the matching relation
is equivalent to the belongs-to relation $\vty \in \interpty{\ty}$. 
Therefore, the definitions~\eqref{eq:semsub-def}
and~\eqref{eq:truesemsub-def-new} are also equivalent.

Since $\bjtruesemsub{\ty_1}{\ty_2}$ is equivalent to $\bjsemsub{\ty_1}{\ty_2}$
and the equivalence relation $\iff$ is transitive,
it suffices to prove the following theorem
to show~\eqref{eq:declsub-correct-truesemsub}.

\begin{theorem}[Correctness of Declarative Subtyping]\label{thm:declsub-correct}
	\[
	\forall \ty_1, \ty_2.\ (\bjsub{\ty_1}{\ty_2} \iff\ \bjsemsub{\ty_1}{\ty_2})
	\]
\end{theorem}

The full proof of~\thmref{thm:declsub-correct} is 
Coq-mechanized~\cite{bib:MiniJlCoq},
so we only discuss some key aspects and leave details to the proof.
First, subtyping a value type coincides with matching:
\begin{equation}\label{eq:mtch-eq-declsub}
\forall \vty, \ty.\ (\bjsub{\vty}{\ty} \iff\ \bjmtch{\vty}{\ty}).
\end{equation}
Having that, we can prove 
$\bjsub{\ty_1}{\ty_2} \implies \bjsemsub{\ty_1}{\ty_2}$,
i.e. the soundness direction of~\thmref{thm:declsub-correct}
(below, we embed the definition~\eqref{eq:semsub-def} of 
matching-based semantic subtyping):
\begin{equation}\label{eq:declsub-imp-semsub}
\forall \ty_1, \ty_2.\ 
(\bjsub{\ty_1}{\ty_2} \implies\ 
\forall \vty.\ [\bjmtch{\vty}{\ty_1} \implies\ \bjmtch{\vty}{\ty_2}]).
\end{equation}
Knowing $\bjsub{\ty_1}{\ty_2}$ and $\bjmtch{\vty}{\ty_1}$,
we need to show that $\bjmtch{\vty}{\ty_2}$.
First, by applying~\eqref{eq:mtch-eq-declsub} to $\bjmtch{\vty}{\ty_1}$,
we get $\bjsub{\vty}{\ty_1}$.
Then, $\bjsub{\vty}{\ty_2}$ follows from $\bjsub{\vty}{\ty_1}$ 
and $\bjsub{\ty_1}{\ty_2}$ by transitivity.
Finally, by applying~\eqref{eq:mtch-eq-declsub} again,
we get $\bjmtch{\vty}{\ty_2}$. \qed

The other direction of~\thmref{thm:declsub-correct} is more challenging: 
\begin{equation}\label{eq:declsub-complete}
\forall \ty_1, \ty_2.\ 
(\bjsemsub{\ty_1}{\ty_2} \implies\ \bjsub{\ty_1}{\ty_2}).
\end{equation}
The key observation here is that~\eqref{eq:declsub-complete} can be shown 
for $\ty_1$ in \emph{normal form},
i.e. $\ty_1 \equiv \vty_1 \cup \vty_2 \cup \ldots \cup \vty_n$
(formally, this fact is denoted by predicate $\InNF(\ty_1)$ 
defined in~\figref{fig:bjsem-innf}, \appref{app:nf}):
\begin{equation}\label{eq:nf-declsub-complete}
\forall \ty_1, \ty_2 \Alt \InNF(\ty_1).\
(\bjsemsub{\ty_1}{\ty_2} \implies\ \bjsub{\ty_1}{\ty_2}).
\end{equation}
In this case, in the definition~\eqref{eq:semsub-def}
of $\bjsemsub{\ty_1}{\ty_2}$, the only
value types $\vty$ that match $\ty_1$ and $\ty_2$ are $\vty_i$ of $\ty_1$. 
By~\eqref{eq:mtch-eq-declsub}, we know that matching implies subtyping,
so we conclude that all $\bjsub{\vty_i}{\ty_2}$.
From the latter, it is easy to show that 
$\bjsub{(\vty_1 \cup \vty_2 \cup \ldots \cup \vty_n)}{\ty_2}$ because,
according to the \RD{UnionL} rule,
subtyping of the left-hand side union amounts to subtyping its components.
To show~\eqref{eq:declsub-complete}, we need several more facts
in addition to~\eqref{eq:nf-declsub-complete}.
\begin{itemize}
  \item Function $\NF$ produces a type in normal form:
	\begin{equation}\label{eq:nf-innf}
	\forall \ty.\ \InNF(\NF(\ty)).
	\end{equation}
  \item Normalized type is equivalent to the source type:
    \begin{equation}\label{eq:nf-declsub-eq}
    \forall \ty.\;\ \bjsub{\NF(\ty)}{\ty}\ \land\ \bjsub{\ty}{\NF(\ty)}.
    \end{equation}
  \item Normalization preserves the subtyping relation:
    \begin{equation}\label{eq:nf-semsub}
    \forall \ty_1, \ty_2.\ 
    (\bjsemsub{\ty_1}{\ty_2} \implies\ \bjsemsub{\NF(\ty_1)}{\ty_2}).
    \end{equation}
\end{itemize}
To prove~\eqref{eq:declsub-complete}, we need to show $\bjsub{\ty_1}{\ty_2}$
given $\bjsemsub{\ty_1}{\ty_2}$. 
For this, we first apply~\eqref{eq:nf-semsub} to $\bjsemsub{\ty_1}{\ty_2}$,
which gives $\bjsemsub{\NF(\ty_1)}{\ty_2}$. 
Then we can apply~\eqref{eq:nf-declsub-complete} to the latter
because of~\eqref{eq:nf-innf} to get $\bjsub{\NF(\ty_1)}{\ty_2}$.
Finally, \eqref{eq:nf-declsub-eq} and transitivity 
gives $\bjsub{\ty_1}{\ty_2}$. \qed

\subsection{Reductive Subtyping}\label{sec:redsub-correct}

Since we have already shown that declarative subtyping is
equivalent to semantic subtyping, it suffices to show that
reductive subtyping is equivalent to declarative subtyping:

\begin{theorem}[Correctness of Reductive Subtyping]\label{thm:redsub-correct}
\[
\forall \ty_1, \ty_2.\ (\bjsubr{\ty_1}{\ty_2} \iff\ \bjsub{\ty_1}{\ty_2})
\]
\end{theorem}

The proof is split into two parts: soundness and completeness.
For soundness (completeness), 
we show that for each \RR{}\!rule (\RD{}\!rule) it is possible to build
a corresponding declarative (reductive) derivation 
using \RD{}\!rules (\RR{}\!rules).

The soundness direction is mostly straightforward, as most \RR{}\!rules
have an immediate \RD{}\!counterpart (or require one extra application
of transitivity).
In the case of \RR{NF}, the induction hypothesis of the proof,
$\bjsub{\NF(\ty_1)}{\ty_2}$, 
and the fact that $\bjsub{\ty_1}{\NF(\ty_1)}$ 
according to~\eqref{eq:nf-declsub-eq},
allow to conclude $\bjsub{\ty_1}{\ty_2}$.

The challenging part of the proof is to show completeness,
as this requires proving that the reductive definition 
is \emph{reflexive}, \emph{transitive}, and \emph{distributive}
(\appref{app:proofs}).


\begin{theorem}[Decidability of Reductive Subtyping]\label{thm:redsub-decidable}
\[
\forall \ty_1, \ty_2.\ 
(\bjsubr{\ty_1}{\ty_2}\quad \lor\quad \lnot [\bjsubr{\ty_1}{\ty_2}])
\]
\end{theorem}

To prove the theorem, 
it suffices to show that reductive subtyping is decidable
when $\ty_1$ is in normal form.
This is done by induction on a derivation of $\InNF(\ty_1)$.
We refer the reader to \appref{app:proofs} for more details.


%% file: discussion.tex
We set out to define semantic subtyping that can be useful in the context
of dynamic languages, however, 
the semantic definition we presented
appears to have an undesired implication for dynamic dispatch.
In this section, using multiple dispatch as a running example,
we discuss the implication and suggest a solution.

Consider the following methods\footnote{In the context of MD, 
	different implementations of 
	the same function are usually called \emph{methods},
	and the set of all methods a \emph{generic function}.}
of the addition function defined in the Julia syntax
(we assume that function \jlcode{flt} converts its argument to a float):
\begin{lstminijl}
+(x::Int, y::Int) = prim_add_int(x, y) 
+(x::Flt, y::Flt) = prim_add_flt(x, y) 
+(x::Int
\end{lstminijl}
and the function call \jlcode{3 + 5}.
With multiple dynamic dispatch, the call is resolved at run-time,
based on the types of all arguments. 
But how exactly does method resolution work?

One approach to implementing multiple dispatch, adopted by some languages
such as Julia~\cite{Bezanson2015AbstractionIT}, 
is to use subtyping on tuple types~\cite{bib:Leavens:1998:mddtuples}.
Namely, method signatures and function calls are interpreted as tuple types,
and then subtyping is used to determine applicable methods 
as well as pick one of them.
In the example above, the three methods are interpreted 
as the following types (from top to bottom):\\
\jlcode{mII $\equiv$ Int $\times$ Int}\\
\jlcode{mFF $\equiv$ Flt $\times$ Flt}\\
\jlcode{mUU $\equiv$ (Int$\cup$Flt) $\times$ (Int$\cup$Flt)}\\
and the call as having type \jlcode{cII $\equiv$ Int $\times$ Int}.
To resolve the call, the language run-time ought to perform two steps.
\begin{enumerate}
  \item Find the applicable methods (or raise an error if there are none). 
    For this, subtyping is checked between
    the type of the call \jlcode{cII} and the method signatures.
    Since \jlcode{cII $<:$ mII} and \jlcode{cII $<:$ mUU} but
    \jlcode{cII $\not{<:}$ mFF}, only two methods are applicable~---
    \jlcode{mII} for integers and \jlcode{mUU} for mixed-type numbers.
  \item Pick the most specific of the applicable methods
    (or raise an error if there is an ambiguity).
    For this, subtyping is checked pairwise between all the applicable methods.
    In this example, naturally, we would like \jlcode{mII} to be called
    for \jlcode{3 + 5}. And indeed, since \jlcode{mII $<:$ mUU}
    and \jlcode{mUU $\not{<:}$ mII},
    the integer addition is picked as the most specific.
\end{enumerate}
As another example, consider the call \jlcode{3.14 + 5}, which type is
\jlcode{Flt $\times$ Int}. There is only one applicable method \jlcode{mUU}
that is a supertype of the call type, so it should be picked.

What happens if the programmer defines
several implementations with the same argument types? 
In the case of a static language, an error can be reported.
In the case of a dynamic language, however, the second implementation
simply replaces the earlier one in the same way as reassignment
to a variable replaces its previous value.

For instance, consider a program 
that contains the three previous implementations of \jlcode{(+)} 
and also: 
\begin{lstminijl}
+(x::Real, y::Real) = ...   # mRR
print(3.14 + 5)	 
\end{lstminijl}
According to the semantic subtyping relation, type \jltype{Real} is equivalent
to \jltype{Int$\cup$Flt} in \BetaJulia. 
Therefore, the implementation of \jlcode{mRR} will replace 
\jlcode{mUU} defined earlier, 
and the mixed-type call \jlcode{3.14 + 5} will be dispatched to \jlcode{mRR}.

But there is a problem: 
the semantics of the program above will change
if the programmer adds a new subtype of \jltype{Real} 
into the nominal hierarchy, e.g. \jlcode{Int8 $<:$ Real}.
In this case, type \jlcode{Real}
stops being equivalent to \jltype{Int$\cup$Flt}
and becomes equivalent to \jltype{Int$\cup$Flt$\cup$Int8}.
Thus, when the program is re-run, type \jlcode{mUU} 
will be a strict subtype of \jlcode{mRR},
so the implementation of \jlcode{mRR} will \emph{not} replace \jlcode{mUU}. 
Therefore, this time, the call \jlcode{3.14 + 5} 
will be dispatched to \jlcode{mUU}, not \jlcode{mRR} as before.

We can gain stability by removing subtyping rules that 
equate abstract nominal types with the union of their subtypes
(i.e. \RD{RealUnion} and \RD{NumUnion} 
in the declarative definition\footnote{To get 
	equivalent reductive subtyping, we need to change 
    the \RR{NF} rule by replacing normalization function $\NF$ with $\NFAt$ 
    (\figref{fig:bjnom-calc-nf-full}, \appref{app:nf}).}
from~\figref{fig:bjsem-decl-sub}).
Then, to fix the discrepancy between the new definition 
and semantic subtyping, the latter should be modified. 
To account for potential extension of the nominal hierarchy,
abstract nominal type \aname can be interpreted 
as containing an extra element $E_\aname$~--- ``a~future subtype of \aname''.
In the case of \BetaJulia, the new interpretation is as follows:
\[
\begin{array}{rcl}
\interpty{\tyreal} & = & \{ \tyint, \tyflt, E_{\tyreal} \} \\
\interpty{\tynum} & = & \{ \tyint, \tyflt, \tycmplx,
                           E_{\tyreal}, E_{\tynum} \}.
\end{array}
\]
It can be shown that
the modified declarative definition of subtyping
is equivalent to semantic subtyping 
based upon the new interpretation.\footnote{The proof can be found 
	in \texttt{FullAtomicJl} folder of~\cite{bib:MiniJlCoq}.}

%% file: related-work.tex
Semantic subtyping has been studied primarily in the context of
\emph{statically typed} languages with \emph{structural} typing. 
For example, \citet{bib:Hosoya:2003:XDuce} defined 
a semantic type system for XML that incorporates unions, products,
and recursive types, with a subtyping algorithm based on tree 
automata~\cite{bib:Hosoya:2005:XML}.
\citet{bib:Frisch:2008:sem-sub} presented decidable semantic subtyping
for a language with functions, products, and boolean combinators 
(union, intersection, negation); the decision procedure 
for $\ty_1 <: \ty_2$ is based on checking
the emptiness of $\ty_1 \setminus \ty_2$. 
\citet{bib:Dardha:2013:semsub-oo} adopted semantic subtyping
to objects with structural types, and \citet{bib:Ancona:2016:sem-sub-oo} 
proposed decidable semantic subtyping for mutable records.
Unlike these works, we are interested in applying semantic reasoning
to a \emph{dynamic} language with \emph{nominal} types.

Though {multiple dispatch} is more often found in dynamic languages,
there has been research on safe integration of dynamic dispatch into
statically typed languages~\cite{bib:Chambers:1992:Cecil, 
Castagna:1992:COF:141471.141537, bib:Clifton:2000:MultiJava,
Allen:2011:TCM:2076021.2048140,Park:2019:PSM:3302515.3290324}. 
There, subtyping is used for both
static type checking and dynamic method resolution.
In the realm of dynamic languages, \citet{Bezanson2015AbstractionIT} 
employed subtyping for multiple dynamic dispatch in the Julia language.
Julia has a rich language of type annotations 
(including, but not limited to, nominal types, tuples, and unions) 
and a complex subtyping relation~\cite{ZappaNardelli:2018:JSR:3288538.3276483}. 
However, it is not clear whether the subtyping relation is decidable 
or even transitive, and transitivity of subtyping is important
for correct implementation of method resolution.
In this paper, while we work with only a subset of Julia types, 
subtyping is transitive and decidable.

Recently, a framework for building transitive, distributive,
and decidable subtyping of union and intersection types was proposed 
by~\citet{Muehlboeck:2018:EUI:3288538.3276482}.
Our language of types does not have intersection types but features
pair types that distribute over unions in a similar fashion.

Finally, \citet{bib:Chung19} proved that Julia's algorithm for subtyping
tuples, unions, and primitive types (without a nominal hierarchy)
is equivalent to a semantic subtyping model similar to ours.
Combined with our results, this shows that a normalization-based
subtyping algorithm for tuples and unions can be implemented efficiently.

%% file: conclusion.tex
We have presented a decidable relation for subtyping of
nominal types, tuples, and unions.
Our system has the advantages of semantic subtyping, 
such as simple set-theoretic reasoning, 
yet it can be used in the context of dynamically typed languages.
We interpret types in terms of type tags, 
as is typical for dynamic languages,
and provide a decidable syntactic subtyping relation 
equivalent to the subset relation of the interpretations
(aka tag-based semantic subtyping).

We found that the initially proposed subtyping relation, 
if used for dynamic dispatch, 
would make the semantics of dynamically typed programs unstable
due to an interaction of abstract nominal types and unions.
A slightly different semantic interpretation of nominal types 
appeared to fix the issue, 
and we would like to further explore this alternative.

In future work, we plan to extend tag-based semantic subtyping 
to top and bottom types, 
and also invariant type constructors such as 
parametric references $\tyref[\ty]$: 
\[
\begin{array}{rcl}
\ty \in \Type   &::=& \ldots \Alt \tyref[\ty]\\
\vty \in \VType &::=& \ldots \Alt \tyref[\ty]
\end{array}
\]
As usual for invariant constructors, 
we would like types such as $\tyref[\tyint]$
and $\tyref[\tyunion{\tyint}{\tyint}]$ to be equivalent.
However, a naive interpretation of invariant types below
is not well defined because to find all $\ty'$ 
s.t. $\interpty{\ty'} = \interpty{\ty}$, 
we need to already know all the interpretations:
\[
\interpty{\tyref[\ty]} = 
\{ \tyref[\ty'] \Alt \vty \in \interpty{\ty} \iff \vty \in \interpty{\ty'} \}.
\]
Our plan is to introduce an indexed interpretation
\[
\interpty{\tyref[\ty]}_{k+1} = \{ \tyref[\ty'] 
    \Alt \vty \in \interpty{\ty}_k \iff \vty \in \interpty{\ty'}_k \}
\]
and define semantic subtyping as:
\[
\bjtruesemsub{\ty_1}{\ty_2} \quad \defsign \quad
\forall k.\ (\interpty{\ty_1}_k \subseteq \interpty{\ty_2}_k).
\]

%% file: appendix.tex
\section{Normal Forms}\label{app:nf}

\begin{figure}
  \begin{mathpar}
  	\inferrule*[right=NF-ValType]
  	{ }
  	{ \InNF(\vty) }
  	
  	\inferrule*[right=NF-Union]
  	{ \InNF(\ty_1) \\ \InNF(\ty_2) }
  	{ \InNF(\tyunion{\ty_1}{\ty_2}) }
  \end{mathpar}
    \caption{Normal form of types in \BetaJulia}
    \label{fig:bjsem-innf}
\end{figure}

\begin{figure}
  \[
	\begin{array}{rcl}
	\NF: \Type &\rightarrow& \Type \\
	\NF(\cname) &=& \cname \\
	\NF(\tyreal) &=& \tyunion{\tyint}{\tyflt} \\
	\NF(\tynum) &=& \tyunion{\tyunion{\tyint}{\tyflt}}{\tycmplx} \\
	\NF(\typair{\ty_1}{\ty_2}) &=& \unprs(\NF(\ty_1), \, \NF(\ty_2))	\\
	\NF(\tyunion{\ty_1}{\ty_2}) &=& \tyunion{\NF(\ty_1)}{\NF(\ty_2)} \\
	& & \\
	\unprs: \Type\times\Type &\rightarrow& \Type \\
	\unprs(\tyunion{\ty_{11}}{\ty_{12}},\ \ty_2) &=&
	  \tyunion{\unprs(\ty_{11}, \ty_2)}{\unprs(\ty_{12}, \ty_2)} \\
	\unprs(\ty_1,\ \tyunion{\ty_{21}}{\ty_{22}}) &=&
	  \tyunion{\unprs(\ty_1, \ty_{21})}{\unprs(\ty_1, \ty_{22})} \\
	\unprs(\ty_1, \, \ty_2) &=& \typair{\ty_1}{\ty_2}
	\end{array}
  \]
	\caption{Computing normal form of \BetaJulia types}
	\label{fig:bjsem-calc-nf-full}
\end{figure}

\begin{figure}
	\begin{mathpar}
		\inferrule[Atom-CName]
		{ }
		{ \Atom(\cname) }
		
		\inferrule[Atom-AName]
		{ }
		{ \Atom(\aname) }
		\\
		
		\inferrule*[right=NFAt-Atom]
		{ \Atom(\ty) }
		{ \InNFAt(\ty) }
		
		\inferrule*[right=AtNF-Union]
		{ \InNFAt(\ty_1) \\ \InNFAt(\ty_2) }
		{ \InNFAt(\tyunion{\ty_1}{\ty_2}) }
	\end{mathpar}
	\caption{Atomic normal form of types in \BetaJulia}
	\label{fig:bjnom-innf}
\end{figure}

\begin{figure}
	\[
	\begin{array}{rcl}
	\NFAt: \Type &\rightarrow& \Type \\
	\NFAt(\cname) &=& \cname \\
	\NFAt(\aname) &=& \aname \\
	\NFAt(\typair{\ty_1}{\ty_2}) &=& \unprs(\NFAt(\ty_1), \, \NFAt(\ty_2))	\\
	\NFAt(\tyunion{\ty_1}{\ty_2}) &=& \tyunion{\NFAt(\ty_1)}{\NFAt(\ty_2)} \\
	\end{array}
	\]
	\caption{Computing atomic normal form of \BetaJulia types}
	\label{fig:bjnom-calc-nf-full}
\end{figure}

\figref{fig:bjsem-innf} defines the predicate $\InNF(\ty)$, which states
that type $\ty$ is in normal form.
\figref{fig:bjsem-calc-nf-full} contains the full definition of $\NF(\ty)$ 
function, which computes the normal form of a type.

\figref{fig:bjnom-innf} and \figref{fig:bjnom-calc-nf-full} present 
``atomic normal form'', which can be used to define reductive subtyping
that disables derivations such as $\bjsub{\tyreal}{\tyunion{\tyint}{\tyflt}}$.

\section{Non-unique Derivations}\label{app:example-deriv}

There are two derivations of
\[\bjsubr{\typair{\tystr}{(\tyunion{\tyint}{\tyflt})}}
         {\typair{\tystr}{\tyreal}}.\]
The shortest derivation:
\begin{mathpar}
\footnotesize
\inferrule*[right=\footnotesize{Pair}]
{ \inferrule[\footnotesize{BaseRefl}]
  { }
  { \bjsubr{\tystr}{\tystr} } \\
  \inferrule*[right=\footnotesize{UnionL}]
  { \inferrule[\footnotesize{IntReal}]
  	{ }{ \bjsubr{\tyint}{\tyreal} } \\
    \inferrule[\footnotesize{FltReal}]
    { }{ \bjsubr{\tyflt}{\tyreal} } }
  { \bjsubr{\tyunion{\tyint}{\tyflt}}{\tyreal} } }
{ \bjsubr{\typair{\tystr}{(\tyunion{\tyint}{\tyflt})}}
	{\typair{\tystr}{\tyreal}} }
\end{mathpar}
The normalization-based derivation:
\begin{mathpar}
\footnotesize
\inferrule*[right=\scriptsize{NF}]
{ \inferrule*[right=\scriptsize{UnionL}]
  { \scriptsize
  	\inferrule
  	{ \inferrule
      { }{ \bjsubr{\tystr}{\tystr} } \\
      \inferrule
      { }{ \bjsubr{\tyint}{\tyreal} } }
  	{ \bjsubr{\typair{\tystr}{\tyint}}{\typair{\tystr}{\tyreal}} } \\
    \scriptsize
    \inferrule
    { \inferrule{ }{\bjsubr{\tystr}{..} } \\
      \inferrule{ }{\bjsubr{\tyflt}{..} } }
    { \bjsubr{\typair{\tystr}{\tyflt}}{\typair{\tystr}{\tyreal}} } }
  { \bjsubr{\tyunion{(\typair{\tystr}{\tyint})}{(\typair{\tystr}{\tyflt})}}
  	  {\typair{\tystr}{\tyreal}} } }
{ \bjsubr{\typair{\tystr}{(\tyunion{\tyint}{\tyflt})}}
	{\typair{\tystr}{\tyreal}} }
\end{mathpar}

\section{Overview of Coq Proofs}\label{app:proofs}

In this section we give a brief overview of 
the Coq-mecha\-ni\-za\-tion~\cite{bib:MiniJlCoq} of the paper.
When referring to a file \fn{fname}, 
we mean the file \fn{Mechanization/fname} in~\cite{bib:MiniJlCoq}.

\subsection{Definitions}

Most of the relevant definitions are in \fn{MiniJl/BaseDefs.v}.
In the table below, 
we show the correspondence between paper definitions (left column)
and Coq definitions (middle column),
possibly with syntactic sugar (right column).

\begin{center}
\begin{tabular}{r|l|l}
\hline
\multicolumn{3}{l}{Types} \\
\hline
\ty  & \coqcode{ty} & \\
\vty & \coqcode{value\_type v} & \\

\hline
\multicolumn{3}{l}{Relations} \\
\hline
$\bjmtch{\vty}{\ty}$ & 
	\coqcode{match\_ty v t} & \coqcode{|- v <\$ t} \\
$\bjsemsub{\ty_1}{\ty_2}$ &
	\coqcode{sem\_sub t1 t2} & \coqcode{||- [t1] <= [t2]} \\
$\bjsub{\ty_1}{\ty_2}$ &
	\coqcode{sub\_d t1 t2} & \coqcode{|- t1 << t2} \\
$\bjsubr{\ty_1}{\ty_2}$ &
	\coqcode{sub\_r t1 t2} & \coqcode{|- t1 << t2} \\
	
\hline
\multicolumn{3}{l}{Auxiliary definitions} \\
\hline
$\InNF(\ty)$ & \coqcode{in\_nf t} & \coqcode{InNF(t)} \\
$\NF(\ty)$   & \coqcode{mk\_nf t} & \coqcode{MkNF(t)} \\
$\unprs(\ty_1, \ty_2)$ &
	\coqcode{unite\_pairs t1 t2} & \\
\hline
\end{tabular}
\end{center}

\subsection{Basic Properties of Normalization Function}

File \fn{MiniJl/BaseProps.v} contains several simple properties 
that are needed for proving the major theorems discussed in the paper,
in particular, the following properties of the normalization function $\NF$:

\begin{center}
\begin{tabular}{r|c|l}
\hline
Statement & Ref in text & Name in Coq \\
\hline
$\InNF(\NF(\ty))$ & \eqref{eq:nf-innf} & \coqcode{mk\_nf\_\_in\_nf} \\
$\InNF(\ty) \implies (\NF(\ty) \equiv \ty)$ &
	& \coqcode{mk\_nf\_nf\_\_equal} \\
$\NF(\NF(\ty)) \equiv \NF(\ty)$ &
    & \coqcode{mk\_nf\_\_idempotent} \\
\hline
\end{tabular}
\end{center}

\subsection{Basic Properties of Matching Relation}

The following properties are proven in \fn{MiniJl/PropsMatch.v}.

\begin{itemize}
  \item Matching relation is \emph{reflexive}, \coqcode{match\_valty\_\_rflxv}
    (by induction on \vty):
    \[\forall \vty.\ \bjmtch{\vty}{\vty}.\]
  \item The only value type that a value type matches is the value type itself,
    \coqcode{valty\_match\_valty\_\_equal} 
    (by induction on $\bjmtch{\vty_1}{\vty_2}$):
    \[\forall \vty_1, \vty_2.\ (\bjmtch{\vty_1}{\vty_2} \implies 
    \vty_1 \equiv \vty_2).\]
  \item The matching relation is \emph{decidable}, 
    \coqcode{match\_ty\_\_dcdbl}
    (by induction on \vty, then by induction on \ty):
    \[\forall \vty, \ty.\ 
    (\bjmtch{\vty}{\ty}\; \lor\; \lnot [\bjmtch{\vty}{\ty}]).\]
\end{itemize}

\subsection{Correctness of Declarative Subtyping}

First, we discuss some auxiliary statements 
that are needed for proving~\thmref{thm:declsub-correct}
(located in \fn{MiniJl/DeclSubProp.v}).

One direction of~\eqref{eq:mtch-eq-declsub},
\begin{equation}\label{eq:mtch-imp-declsub}
\forall \vty, \ty.\ (\bjmtch{\vty}{\ty} \implies\ \bjsub{\vty}{\ty}),
\end{equation}
is proven in \coqcode{match\_ty\_\_sub\_d\_sound} 
by induction on $\bjmtch{\vty}{\ty}$.
The other direction,
\[
\forall \vty, \ty.\ (\bjsub{\vty}{\ty} \implies\ \bjmtch{\vty}{\ty}),
\]
is proven in \coqcode{match\_valty\_\_sub\_d\_complete}
by induction on $\bjsub{\vty}{\ty}$.
The transitivity case, \RD{Trans}, requires a helper statement,
\coqcode{match\_valty\_\_transitive\_on\_sub\_d}:
\begin{equation}\label{eq:mtch-trans}
\forall \ty_1, \ty_2, \vty.\ 
(\bjsub{\ty_1}{\ty_2}\; \land\; \bjmtch{\vty}{\ty_1} \quad \implies \quad
\bjmtch{\vty}{\ty_2}),
\end{equation}
which is proven by induction on $\bjsub{\ty_1}{\ty_2}$.

The equivalence of a type and its normal form~\eqref{eq:nf-declsub-eq}
is shown by induction on \ty
in lemmas \coqcode{mk\_nf\_\_sub\_d1} ($\bjsub{\NF(\ty)}{\ty}$)
and \coqcode{mk\_nf\_\_sub\_d2} ($\bjsub{\ty}{\NF(\ty)}$).

Semantic completeness of declarative subtyping 
for a normalized type~\eqref{eq:nf-declsub-complete},
\[
\forall \ty_1, \ty_2 \Alt \InNF(\ty_1).\
(\bjsemsub{\ty_1}{\ty_2} \implies\ \bjsub{\ty_1}{\ty_2}),
\]
is shown in \coqcode{nf\_sem\_sub\_\_sub\_d} by induction on $\InNF(\ty_1)$.
When $\ty_1 \equiv \vty$, we use~\eqref{eq:mtch-imp-declsub}.
By definition of $\bjsemsub{\vty}{\ty_2}$, we know that 
$\bjmtch{\vty}{\ty_2}$ follows from $\bjmtch{\vty}{\vty}$.\\
When $\ty_1 \equiv \tyunion{\ty_a}{\ty_b}$, 
we use induction hypothesis $\bjsub{\ty_a}{\ty_2}$ and $\bjsub{\ty_b}{\ty_2}$,
\RD{UnionL} rule, and the fact that
\[\forall \vty, \ty_1, \ty_2.\ 
(\bjmtch{\vty}{\ty_i} \implies \bjmtch{\vty}{\tyunion{\ty_1}{\ty_2}}). \]

Finally, soundness and completeness parts of~\thmref{thm:declsub-correct}
(\coqcode{sub\_d\_\_semantic\_sound} and \coqcode{sub\_d\_\_semantic\_complete})
are proven in \fn{MiniJl/Props.v}.
Note that soundness~\eqref{eq:declsub-imp-semsub} is the same as
transitivity of the matching relation~\eqref{eq:mtch-trans}.
The completeness part~\eqref{eq:declsub-complete} is proven as explained
at the end of \secref{sec:declsub-correct}.

\subsection{Correctness of Reductive Subtyping}

As discussed in~\secref{sec:redsub-correct}, the soundness part
of \thmref{thm:redsub-correct} 
(lemma \coqcode{sub\_r\_\_sound} in \fn{MiniJl/Props.v}),
\[
\forall \ty_1, \ty_2.\ (\bjsubr{\ty_1}{\ty_2} \implies\ \bjsub{\ty_1}{\ty_2}),
\]
is proven by induction on $\bjsubr{\ty_1}{\ty_2}$.
The only interesting case is the rule \RR{NF} where we have
the induction hypothesis $\bjsub{\NF(\ty_1)}{\ty_2}$ 
and need to show $\bjsub{\ty_1}{\ty_2}$.
Since $\bjsub{\ty_1}{\NF(\ty_1)}$, we can use transitivity (rule \RD{Trans}).

The completeness part of \thmref{thm:redsub-correct}
(lemma \coqcode{sub\_r\_\_complete} in \fn{MiniJl/Props.v}),
\[
\forall \ty_1, \ty_2.\ (\bjsub{\ty_1}{\ty_2} \implies\ \bjsubr{\ty_1}{\ty_2}),
\]
is ultimately proven by induction on $\bjsub{\ty_1}{\ty_2}$.
However, the proof requires showing that reductive subtyping
satisfies the following properties (defined in \fn{MiniJl/RedSubProps.v}):
\begin{itemize}
  \item \emph{Reflexivity}, \coqcode{sub\_r\_\_reflexive} 
    (by induction on \ty):
    \[\forall \ty.\ \bjsubr{\ty}{\ty}. \]
  \item \emph{Transitivity}, \coqcode{sub\_r\_\_transitive}:
    \[
    \forall \ty_1, \ty_2, \ty_3.\ 
    (\bjsubr{\ty_1}{\ty_2}\ \land\ \bjsubr{\ty_2}{\ty_3} \; \implies \;
    \bjsubr{\ty_1}{\ty_3}).
    \]
  \item \emph{Distributivity} of pairs over unions:
    \[
    \bjsubr{\typair{(\tyunion{\ty_{11}}{\ty_{12}})}{\ty_2}}
    {\tyunion{(\typair{\ty_{11}}{\ty_2})}{(\typair{\ty_{12}}{\ty_2})}}
    \]
    and
    \[
    \bjsubr{\typair{\ty_1}{(\tyunion{\ty_{21}}{\ty_{22}})}}
    {\tyunion{(\typair{\ty_1}{\ty_{21}})}{(\typair{\ty_1}{\ty_{22}})}}.
    \]
\end{itemize}

The transitivity proof is done by induction on $\bjsubr{\ty_1}{\ty_2}$.
In some cases it relies on the fact that subtyping a type is the same
as subtyping its normal form,
\begin{equation}\label{eq:redsub-imp-nfredsub1}
\forall \ty.\ (\bjsubr{\ty_1}{\ty_2} \; \iff \; \bjsubr{\NF(\ty_1)}{\ty_2}).
\end{equation}
The right-to-left part follows from \RR{NF}, 
and the left-to-right is shown by induction on $\bjsubr{\ty_1}{\ty_2}$ 
(\coqcode{sub\_r\_\_mk\_nf\_sub\_r1}).
In the \RR{Pair} case of the transitivity proof,
we also need to perform induction on $\bjsubr{\ty_2}{\ty_3}$.
The last case, \RR{NF}, uses the two auxiliary facts:
\[
\forall \ty_1, \ty_2.\ 
(\bjsubr{\ty_1}{\ty_2}\; \implies\; \bjsubr{\NF(\ty_1)}{\NF(\ty_2)}),
\]
proven in \coqcode{sub\_r\_\_mk\_nf\_sub\_r} 
by induction on $\bjsubr{\ty_1}{\ty_2}$
(uses the idempotence of $\NF$),
and $\forall \ty_1, \ty_2, \ty_3.$
\[
\InNF(\ty_1) \land \InNF(\ty_2) \land
(\bjsubr{\ty_1}{\ty_2}) \land (\bjsubr{\ty_2}{\ty_3}) \ \implies \
\bjsubr{\ty_1}{\ty_3},
\]
proven in \coqcode{sub\_r\_nf\_\_transitive} 
by induction on $\bjsubr{\ty_1}{\ty_2}$.

The distributivity proofs use the fact that
\[
\forall \ty_1, \ty_2.\ 
(\bjsubr{\NF(\ty_1)}{\NF(\ty_2)}\; \implies\; \bjsubr{\ty_1}{\ty_2}),
\]
proven in \coqcode{mk\_nf\_sub\_r\_\_sub\_r},
and that normal forms of both types in \RD{Distr*} rules
are in the subtyping relation:
\[
\bjsubr{\NF(\typair{(\tyunion{\ty_{11}}{\ty_{12}})}{\ty_2})}
{\NF(\tyunion{(\typair{\ty_{11}}{\ty_2})}{(\typair{\ty_{12}}{\ty_2})})}
\]
(\coqcode{mk\_nf\_\_distr11}) and
\[
\bjsubr{\NF(\typair{\ty_1}{(\tyunion{\ty_{21}}{\ty_{22}})})}
{\NF(\tyunion{(\typair{\ty_1}{\ty_{21}})}{(\typair{\ty_1}{\ty_{22}})})}
\]
(\coqcode{mk\_nf\_\_distr21}).

\subsection{Decidability of Reductive Subtyping}

The proof of \thmref{thm:redsub-decidable},
\[
\forall \ty_1, \ty_2.\ 
(\bjsubr{\ty_1}{\ty_2}\quad \lor\quad \lnot [\bjsubr{\ty_1}{\ty_2}]),
\]
is given by \coqcode{sub\_r\_\_decidable} in \fn{MiniJl/Props.v}.
It relies on the fact (discussed below) 
that reductive subtyping is decidable for $\ty_1$ s.t. $\InNF(\ty_1)$.
\begin{itemize}
  \item Namely, if $\bjsubr{\NF(\ty_1)}{\ty_2}$, 
	then $\bjsubr{\ty_1}{\ty_2}$ by \RR{NF}.
  \item Otherwise, if $\lnot[\bjsubr{\NF(\ty_1)}{\ty_2}]$, 
    which in Coq means
    $\bjsubr{\NF(\ty_1)}{\ty_2} \implies \False$, 
    we can show $\lnot [\bjsubr{\ty_1}{\ty_2}]$ 
    by assuming that $\bjsubr{\ty_1}{\ty_2}$,
    applying~\eqref{eq:redsub-imp-nfredsub1} to it, 
    and thus getting contradiction.
\end{itemize}
Decidability of subtyping of a normalized type,
\[
\forall \ty_1, \ty_2 \Alt \InNF(\ty_1).\ 
(\bjsubr{\ty_1}{\ty_2}\quad \lor\quad \lnot [\bjsubr{\ty_1}{\ty_2}]),
\]
(lemma \coqcode{nf\_sub\_r\_\_decidable} in \fn{MiniJl/RedSubProps.v})
is pro\-ven by induction on $\InNF(\ty_1)$ 
and uses the decidability of the matching relation,
which coincides with reductive subtyping on a value type.